\documentclass[]{aanew}
\usepackage{psfig,graphicx}
\usepackage[authoryear]{natbib}
\bibpunct{(}{)}{;}{a}{}{,}
\begin{document}

\title{A WN4 companion to \object{BD +62\degr2296} in Cas OB5
\thanks{Based on observations made at 
Observatoire de Haute Provence (CNRS), France, and the Isaac Newton
Telescope (La Palma, Spain)}}
\headnote{Research Note}
\author{I.~Negueruela\inst{1,2}}                   
                                                            
\institute{
Dpto. de F\'{\i}sica, Ing. de Sistemas y Teor\'{\i}a de
la Se\~{n}al, Universidad de Alicante, Apdo. 99, E03080 Alicante,
Spain
\and
Observatoire de Strasbourg, 11 rue de l'Universit\'{e},
F67000 Strasbourg, France
}

\mail{ignacio@dfists.ua.es}

\date{Received    / Accepted     }

\titlerunning{Wolf-Rayet companion to \object{BD +62\degr2296}}

\abstract{
I report observations of the triple system \object{BD
+62\degr2296} showing that all its components are early-type stars,
most likely physically related. The faintest component \object{BD
+62\degr2296B} is a hitherto uncatalogued Wolf-Rayet star.
The brightest component, star A, is shown to
be a seemingly normal B2.5Ia supergiant. Long-slit 
spectroscopy of \object{BD +62\degr2296B} shows it to be a narrow-lined 
WN4 star. Given the spatial separation, the two objects are
unlikely to form a physical binary. Spectra of the third
visual component, \object{BD +62\degr2296C}, allow its classification
as a B0III star. Such concentration of massive stars strongly suggests
that \object{BD +62\degr2296} is in reality a very compact 
young open cluster in the area of the OB association Cas OB5. 
}

\maketitle 


\section{Introduction}

During a recent survey in search of distant OB stars \citep{main},
intermediate-resolution spectroscopy was obtained with the {\em
Aur\'{e}lie} spectrograph on the 1.52-m telescope at the Observatoire
de Haute Provence (OHP) for a
number of luminous stars in the region of the association \object{Cas
OB5}.
As part of this programme, a classification spectrum of
\object{BD +62\degr2296A} was taken.
On site examination of the raw spectrum of \object{BD +62\degr2296A}
revealed the presence of a relatively strong and very broad emission
line at the wavelength of \ion{He}{ii}~$\lambda4686$\AA, completely
unprecedented at this spectral type. A reference to \object{BD
+62\degr2296B} being a Wolf-Rayet star appears in \citet{bat94}, but
no Wolf-Rayet star is catalogued at this position. As
{\em Aur\'{e}lie} offers no 
spatial resolution and the seeing during the observations was rather
poor, further observations of the components of \object{BD
+62\degr2296} have been taken 
with long-slit spectrographs. These observations confirm that \object{BD
+62\degr2296B} is a Wolf-Rayet star and suggest that \object{BD
+62\degr2296} is a compact cluster of young stars.  

\begin{figure}
\begin{picture}(250,240)
\put(0,0){\includegraphics{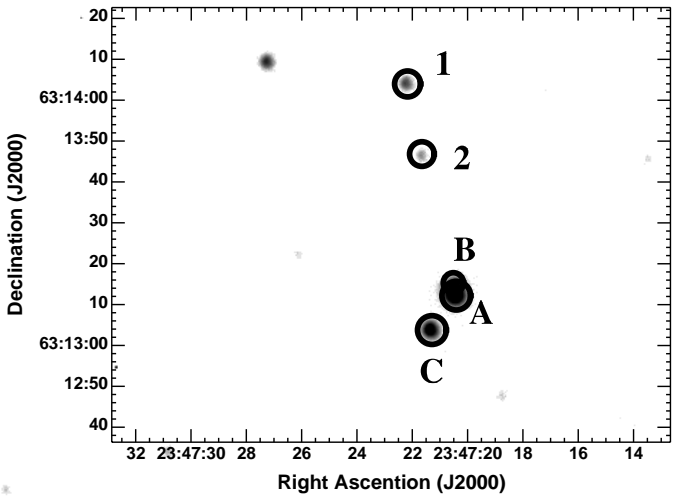}}
\end{picture}
  \caption{Close-up of the area surrounding \object{BD +62\degr2296}
   ($u$-band image taken with the 2.2-m telescope at Calar Alto on
   October 24th, 2002, using the BUSCA camera). The
   components of \object{BD +62\degr2296} are marked as A , B and
   C. Stars \#1 \& \#2  are B-type stars. Positions
   for the components are listed in Table~\ref{tab:pos}.} 
   \label{fig:img}
\end{figure}

In spite of its brightness, \object{BD +62\degr2296} has
not been intensively studied, the only spectral classification for
\object{BD +62\degr2296A} (B3Ia?) 
dating back to \citet{mor55}. \object{BD +62\degr2296A} does not
appear to be photometrically 
very variable: \citet{haug} gives $V=8.64$, $(B-V)=1.07$, while 
\citet{hil56} gives $V=8.65$,
$(B-V)=1.09$. Its radial velocity, averaging $v=-60.3\:{\rm km}\,{\rm
s}^{-1}$, was found to be variable by  
\citet{abt72}. \object{BD +62\degr2296} is catalogued as a visual
triple system \citep{dem83}. \object{BD +62\degr2296C}, also
catalogued as \object{CSI +62\degr2296 2}, is clearly separated,
situated approximately 
$11\arcsec$ to the SE (see Fig.~\ref{fig:img}), while \object{BD
+62\degr2296B} is separated by less than 2\arcsec. With such
small separation, photoelectric measurements must contain both stars. 
The WDS catalogue gives magnitudes $V_{T}=8.93$ and $V_{T}=11.20$ from {\em
Tycho} \citep{fab02} for components A and B. These are untransformed
{\em Tycho} magnitudes (see \citealt{main} for a discussion on the
danger of using transformed magnitudes for reddened early-type
stars).

\section{Observations}
Spectroscopy of \object{BD +62\degr2296A} was obtained with the {\em
Aur\'{e}lie} spectrograph on the 1.52-m telescope at the Observatoire
de Haute Provence (OHP) in January 2002. The spectrograph was equipped
with grating \#3 and the Horizon 2000 EEV CCD camera (see \citealt{main}
for details). This configuration gives a
dispersion of $\sim 0.22$\AA/pixel, covering a wavelength range of $\approx
440$\AA. \object{BD +62\degr2296A} was observed in four spectral
regions, centred on $\lambda=4175$\AA\ (Jan 18), $\lambda=4680$\AA\
(Jan 19), $\lambda=5850$\AA\ and $\lambda=6600$\AA\ (both on Jan 20). 

\begin{table}
 \caption{{\em Tycho} positions
   \citep{fab02} for the three components of \object{BD +62\degr2296}. }
\begin{center}
\begin{tabular}{ccc}
\hline
Component & R.A. (J2000) & Dec. (J2000) \\
\hline
A & 23 47 20.4 & +63 13 12\\ 
B & 23 47 20.4 & +63 13 14\\
C & 23 47 21.3 & +63 13 04\\
\hline
\end{tabular}
\end{center}
 \label{tab:pos}
\end{table}

Further spectroscopy was taken with the {\em Carelec} spectrograph on
the OHP 1.93-m on the nights of 6th and 7th July,  2002. I used
grating \#1 (1200 ln mm$^{-1}$), which gives a nominal dispersion
$\sim 0.45$\AA/pixel in the blue range (resolution element $\sim
1.5$\AA) . During the first night, the central wavelength was set 
to $\lambda4420$\AA. In the second night, two settings, centred at
$\lambda4450$\AA\ and $\lambda5450$\AA\ were used.

Finally, spectroscopy was taken on July 22nd-25th 2002, using the
2.5-m Isaac Newton Telescope (INT) at La Palma, Spain. 
The telescope was equipped with the Intermediate Dispersion
Spectrograph (IDS) and the 235-mm camera. On the night of the 22nd,
the detector used was an EEV10 CCD, while on the three
other nights it was the Tek\#5 CCD.
The slit width was set to $1\farcs2$ and the R1200Y grating was
used. This configuration gives a nominal
dispersion of $\approx 0.48$\AA/pixel with the EEV and
$\approx0.8$\AA/pixel with the Tek\#5. During these nights,
observations were performed with seeing slightly better than
$1\arcsec$. Under these conditions, components A and B were clearly
resolved. Exposures of all three
components were obtained at different wavelengths,
taking care that the other components were left outside the slit. 

\begin{table}
 \caption{Log of spectroscopic observations of the three components of
\object{BD +62\degr2296}. The brighter component A is a B2.5Ia
supergiant; component C is B0III, while component B is
WN4. Observations marked A+B were performed under seeing conditions
not allowing the separation of the two components. All dates refer to
2002.}
\begin{center}
\begin{tabular}{lccccccc}
\hline
Component & Telescope & Date & Wavelength Range\\
\hline
A + B & OHP 1.52-m & Jan 18 & $3960 -4410$\AA \\ 
A + B & OHP 1.52-m & Jan 19 & $4460 - 4910$\AA \\
A + B & OHP 1.52-m & Jan 20 & $5630-6080$\AA\\
A + B & OHP 1.52-m & Jan 20 & $6890-6940$\AA\\
A + B & OHP 1.93-m & Jul 6 & $3960 -4980$\AA \\
C & OHP 1.93-m & Jul 6 & $3960 -4980$\AA \\
A + B & OHP 1.93-m & Jul 7 & $3960 - 5900$\AA \\
A & INT 2.5-m & Jul 22 & $3750 - 8000$\AA \\
C & INT 2.5-m & Jul 22 & $3750 - 5100$\AA \\
B & INT 2.5-m & Jul 23 & $4050 - 4910$\AA \\
C & INT 2.5-m & Jul 24 & $6300 - 7100$\AA \\
B & INT 2.5-m & Jul 25 & $6300 - 7100$\AA \\
\hline
\end{tabular}
\end{center}
 \label{tab:obs}
\end{table}

The complete log of observations is given in Table~\ref{tab:obs}. All
the data have been reduced with the {\em Starlink} 
packages {\sc ccdpack} \citep{draper} and {\sc figaro}
\citep{shortridge} and analyzed using {\sc figaro} and {\sc dipso}
\citep{howarth}. 

\begin{figure*}
\begin{picture}(500,260)
\put(0,0){\includegraphics{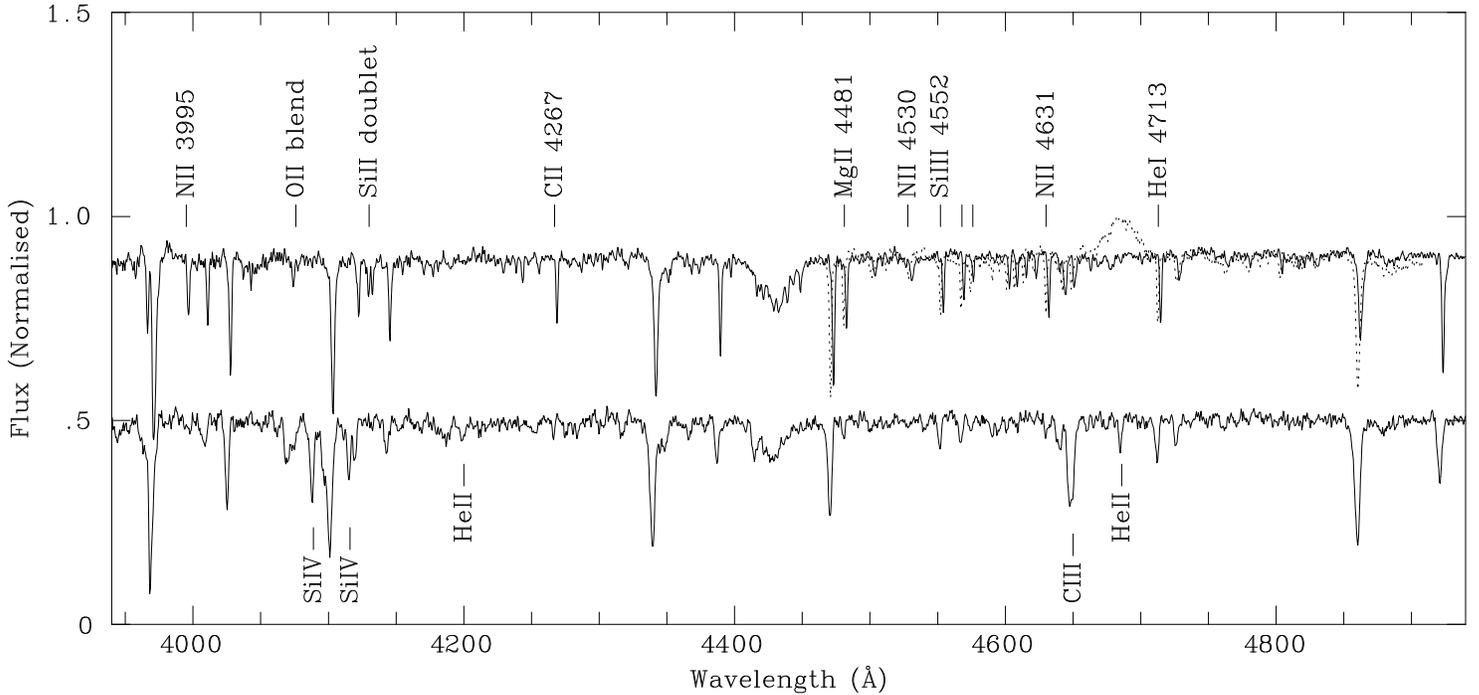}}
\end{picture}
  \caption{Blue spectra of \object{BD +62\degr2296A} (top) and
   \object{BD +62\degr2296C} (bottom), taken on July 22nd, 2002, with
   the INT. The strength of \ion{N}{ii} lines and
   \ion{Mg}{ii}~$\lambda4481$\AA\ in the spectrum of \object{BD
   +62\degr2296A} identify the star as a supergiant. The spectrum of
   \object{BD +62\degr2296A} taken with the 1.52-m OHP on Jan 19th,
   2002, is superimposed as a dotted line. The emission feature at
   $\lambda4686$\AA, completely unexpected at this spectral type,
   arises from the unresolved companion.}
   \label{fig:normals}
\end{figure*}
 
\section{Results}

The spectrum of \object{BD +62\degr2296A} obtained with {\em
Aur\'{e}lie} can be seen in Fig.~\ref{fig:normals}, as a dotted
line. The most striking feature is clearly the prominent broad emission line 
centred on the wavelength of \ion{He}{ii}~$\lambda$4686\AA. This line
is not seen at all at this spectral type, while its presence in
emission is an indication of strong mass loss in bright O-type stars
and Wolf-Rayet stars.

\begin{figure*}
\begin{picture}(500,260)
\put(0,0){\includegraphics{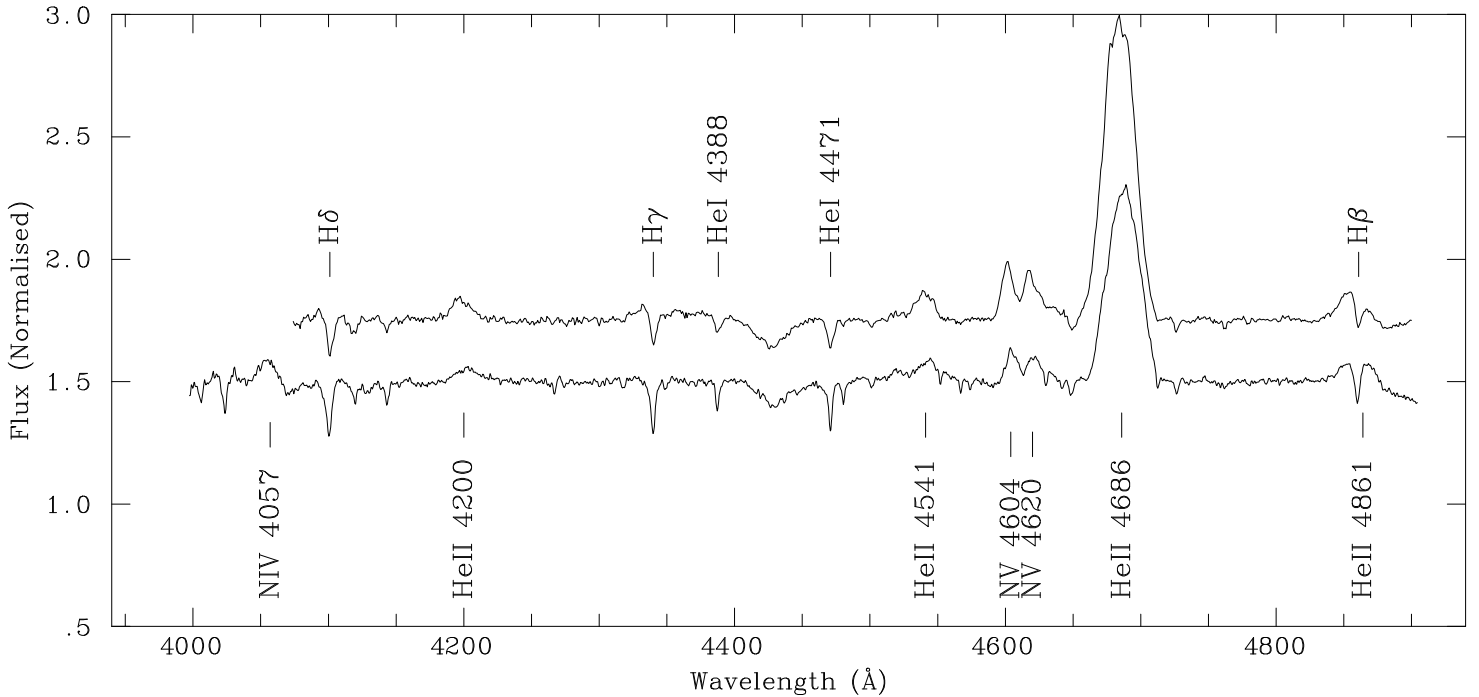}}
\end{picture}
  \caption{Blue spectra of \object{BD +62\degr2296B}. 
	The bottom spectrum was taken on July 7th 2002 with
   the 1.93-m OHP telescope and represents the extraction of a 5-pixel-wide 
   strip centred 4 pixels away from the spectrum peak and therefore
   includes an important contribution from the spectrum of \object{BD
   +62\degr2296B}, which was not resolved in relatively poor seeing. The top
   spectrum was taken on July 25th 2002 with the INT and the slit was
   set on top of the WR star (which was resolved) and perpendicular
   to the axis joining the two stars. Though all emission lines are
	stronger, a small light 
   contribution from the supergiant is evident (note also the lower
	spectral resolution).}
   \label{fig:wr}
\end{figure*}

Unlike {\em Aur\'elie}, the long-slit spectrograph {\em Carelec}
produces spatially resolved spectra.  In the spectrum taken
on July 7th 2002, with a seeing $\approx 2\arcsec$ (estimated from the
FWHM of the spectra of stars \#1 and \#2), components A and B are not
entirely resolved. By
extracting  5 pixels on the Northern side of the spectrum, centred 4
pixels away from the peak in the spatial direction, I obtained a
spectrum of component B, displayed in Fig.~\ref{fig:wr}.

Fig.~\ref{fig:normals} shows a classification spectrum of \object{BD
+62\degr2296A}, taken with the INT, without any contribution from the
WR star. The presence of moderately strong
\ion{Mg}{ii}~$\lambda4481$\AA\ confirms that the star is a mid-range B
supergiant. The
strength of the \ion{O}{ii} spectrum and the relative weakness of
\ion{Mg}{ii}~$\lambda4481$\AA\ support a spectral type around B3Ia. At
B2Ia, \ion{Si}{iii}~$\lambda4552$\AA\ is almost as strong as
\ion{He}{i}~$\lambda4471$\AA\ and \ion{C}{ii}~$\lambda4267$\AA\ is
still weak compared to the \ion{O}{ii} lines. \object{BD +62\degr2296A}
is therefore later and a spectral type B2.5Ia seems adequate. A lower
luminosity class is precluded by the strength of the
\ion{Si}{ii}~$\lambda\lambda4128-30$\AA\ doublet in comparison to the
neighbouring \ion{He}{i} lines (for classification criteria, see
\citealt{waf} and \citealt{len90}).

The INT spectrum of \object{BD +62\degr2296B} is shown in
Fig.~\ref{fig:wr}. It is typical of a WN4
star. \ion{N}{iii}~$\lambda$4640\AA\ is hardly seen, while 
\ion{N}{iv}~$\lambda$4057\AA\ and \ion{N}{v}~$\lambda$4604\AA\ have
comparable strengths. All the other line ratios are compatible with
this spectral type. The line widths are typical of a narrow-lined
Wolf-Rayet. 
The red spectrum of \object{BD +62\degr2296B} taken on July 25th
displays broad strong emission lines corresponding to
\ion{He}{ii}~$\lambda$6560\AA\ and the
\ion{N}{iv}~$\lambda$7103$-$7128\AA\ complex (the latter just on the
spectrum edge). Very weak lines
corresponding to \ion{He}{ii}~$\lambda\lambda$6406 \& 6683 \AA\ are
likely present, as well. The presence of all these lines is in perfect
accord with the spectral classification derived from the blue. Both
the blue and red spectra show features attributable to component A, in
spite of the orientation of the slit intending to leave it out.

The only spectrum including \object{BD +62\degr2296B} in the yellow
region is that of July 7th. Unfortunately, it is less well resolved
than the blue one from the same date, 
perhaps reflecting the larger difference in magnitude between the
supergiant and the Wolf-Rayet star towards longer wavelengths. In
spite of this, broad \ion{He}{ii}~$\lambda$5411\AA\ and
\ion{C}{iv}~$\lambda$5808\AA\ emission lines are seen, with comparable
strength. \ion{He}{i}~$\lambda$5875\AA\ is weakly in emission, though
its strength is difficult to assess due to its nearness to the edge of
the spectrum. All these lines must be contributed by the WN4 star.

\section{Discussion}

\begin{figure}
\begin{picture}(250,260)
\put(0,0){\includegraphics{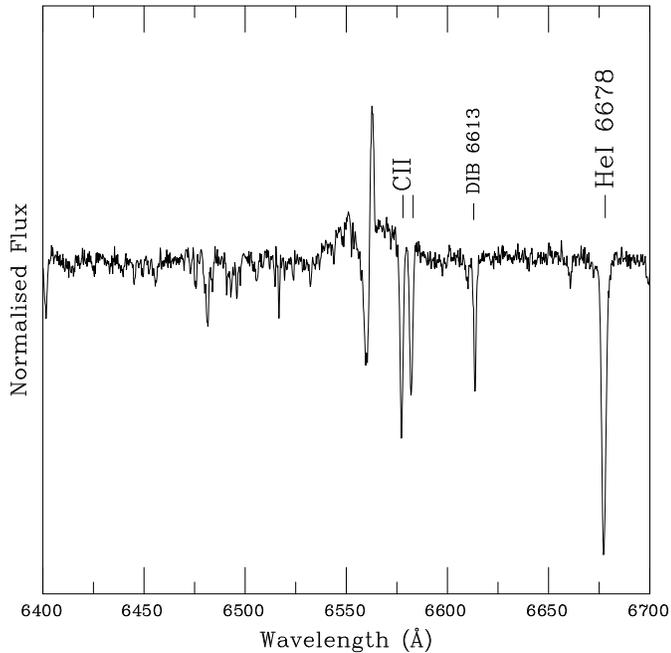}}
\end{picture}
  \caption{H$\alpha$ spectrum of \object{BD +62\degr2296A}, taken on
   Jan 20, 2002. Stellar features are marked, except for H$\alpha$,
   which clearly has two components, a narrow P-Cygni profile
   superimposed on a much broader base.}
   \label{fig:red}
\end{figure}

The close companion to the B2.5Ia supergiant \object{BD
+62\degr2296A},  \object{BD +62\degr2296B}, is
a WN4 Wolf-Rayet star. Its Wolf-Rayet number will be \object{WR\,159}
(van der Hucht, priv. comm.) \object{BD +62\degr2296} appears listed as an
emission line star in the catalogue of \citet{wack}.
The H$\alpha$ spectrum of \object{BD +62\degr2296A} is displayed
in Fig.~\ref{fig:red}. There is a narrow P-Cygni profile, with an edge
velocity of $v_{\rm edge} \approx 500\:{\rm km}\,{\rm s}^{-1}$,
superimposed on a broad emission bump. 
Such features are typical of blue
supergiants of high luminosity and indicate the presence of a
relatively slow radiative wind. From comparison of the OHP and
INT spectra, there does not seem to be a significant contribution from
the WR star to this emission feature. 

From its spectrum (see Fig.~\ref{fig:normals}), \object{BD
+62\degr2296C} is a B0III star. Its $V_{T}$ magnitude is 10.74 from {\em
Tycho}. The difference in magnitude
with component A is compatible with their being at the same distance.
For \object{BD +62\degr2296A}, assuming the intrinsic colour of a
B2.5Ia star $(B-V)_{0}=-0.14$ 
\citep{weg94}, a dereddened distance modulus $DM=12.0$ for \object{Cas
OB5} \citep{humphreys} and a standard reddening law ($R=3.1$), we 
find $M_{V}=-7.1$ using the photometry from
\citet{haug} (though this measurement very likely includes
component B). This
luminosity is typical of a bright supergiant of this spectral
type.

The absorption lines in the spectrum of \object{BD +62\degr2296B}
appear to be due to light contamination from component A (for example,
the characteristic \ion{C}{ii} doublet is seen in absorption close to
\ion{He}{ii}~$\lambda$6560\AA). This does
not rule out the possibility that component C is itself a close
binary. As a matter of fact, if it is at the same distance as
component A, and  its
{\em Tycho} magnitude is correct, it would
be rather more luminous than corresponds to its spectral type. The average
absolute magnitude of a WN4 star is $M_{V}=-3.5$ \citep{hucht}.

 At a distance in
excess of 2 kpc, an angular separation of $\approx 
2\arcsec$ implies a distance 
between components A and B ($\approx 5000\:$AU) far too large to be
consistent with their being an interacting binary. The composite
spectrum shown by \object{BD +62\degr2296A} under 
moderate seeing serves as a cautionary tale when considering
observations of more distant massive stars. Only because the system is
relatively nearby are we able to resolve components A and B, in spite
of their rather large spatial separation.

The fact that the three components of \object{BD +62\degr2296} are
early-type massive stars can hardly be due to a chance alignment. As a
matter of fact, two other objects which fell along the slit (marked \#1
\& \#2 in Fig.~\ref{fig:img}) are also B-type stars.  \object{BD
+62\degr2296} 
is therefore likely a previously unrecognised compact young open
cluster. A photometric study of this area will be presented in a
future paper.

\begin{acknowledgements}

The author would like to thank all the staff at the Observatoire de
Haute Provence for their support and friendliness.  The INT is
operated on the island of La 
Palma by the Isaac Newton Group in the Spanish Observatorio del Roque
de Los Muchachos of the Instituto de Astrof\'{\i}sica de Canarias. The
July 22nd observations were obtained as part of the ING service scheme.

The author would very much like to thank Dr. Karel van der Hucht for
his many important remarks.
The image shown in Figure~1 was kindly provided by Dr. Amparo Marco.  
Many thanks to Simon Clark, Paul Crowther, Artemio Herrero and Miguel
Urbaneja for their comments.  

This research has made use
of the  Simbad data base, operated at CDS,
Strasbourg, France. This research has made use of the Washington
Double Star Catalog maintained at the U.S. Naval Observatory. 

 During part of
this work,  IN has been a researcher of the
programme {\em Ram\'on y Cajal}, funded by the Spanish Ministerio de
Ciencia y Tecnolog\'{\i}a and the University of Alicante.
This research is partially supported by the Spanish Ministerio de
Ciencia y Tecnolog\'{\i}a under grant AYA2002-00814.

\end{acknowledgements}

\end{document}